\documentstyle[aps,epsfig,multicol,eqsecnum,amsfonts]{revtex}

\begin{document}

\title{Asymptotic Distribution of Eigenvalues for a Self-Affine String}
\author{Ingve Simonsen\footnote{E-mail: Ingve.Simonsen@phys.ntnu.no}}
\address{Department of Physics, \\
The Norwegian University of Science and Technology, \\
N--7491 Trondheim, Norway}
\author{Alex Hansen\footnote{Permanent address: Department of Physics, 
The Norwegian University of Science and Technology, NTNU, 
N--7491 Trondheim, Norway. E-mail: Alex.Hansen@phys.ntnu.no}}
\address{International Centre for Condensed Matter Physics, \\
University of Bras{\'\i}lia, \\
CP 04513, 70919--970 Bras{\'\i}lia, Brazil}

\date{\today}

\maketitle
%----------------------------------------------------------------------
%  ABSTRACT
%----------------------------------------------------------------------
\begin{abstract}
We consider a string with fixed endpoints where the mass density and/or
the elastic coefficient vary in a self-affine way as function of position.
It is demonstrated how the eigenvalues in the asymptotic limit are
distributed. Scaling laws for the Weyl term of the asymptotic
integrated density of states is established and confirmed numerically.    
\end{abstract}
\vspace{5mm}
\pacs{PACS numbers: }
% --------------------------------------------------------------------
%   MAIN TEXT
% --------------------------------------------------------------------
\begin{multicols}{2}
\narrowtext

\section{Introduction}

An old problem in mathematical physics is how various irregularities
influence the asymptotics of the cumulative eigenvalue distribution
for a physical resonator.  This problem has many significant physical
applications, such as wave scattering from fractal surfaces, liquid
flow in porous media, vibrations of cracked bodies or macro-molecules
(polymers) etc.

In 1910 Lorentz put forward the conjecture that the number of
eigenmodes up to some (large) frequency $\omega$, depends on the
``volume'' and not on the shape of the resonator. This was later,
proved by Weyl, under condition of a sufficiently smooth but otherwise
arbitrary boundaries \cite{Weyl}.  Later Hunt et al.\ \cite{Hunt}
improved this formula by including a correction term which depends on
lower powers of the frequency and also on the ``surface area'' of the
resonator's perimeter.

Berry \cite{Berry}, when working on wave scattering from fractal
surfaces, generalized the result of Hunt et al. to fractal boundaries.
He conjectured that for fractal boundaries with Hausdorff dimension
$D_H$, the first correction term should scale as $\omega^{D_H}$.
However, Lapidus has proved \cite{Lapidus0} that the correct fractal
dimension is not the Hausdorff dimension, but another nontrivial
dimension known as the Minkowski dimension $D_M$ \cite{Falconer}.

Over the last decade or so, there have been renewed interest in this,
and related problems, both in the mathematics
\cite{Fleckinger2,Lapidus0,Lapidus1,Fleckinger1,Lapidus2,Lapidus3,Lapidus4}
and physics \cite{Sapoval1,Sapoval2,Hobiki1,Hobiki2} communities.  In
this paper we consider a related problem where we study a string which
has a irregular (self-affine) mass density and local elastic
coefficient.  For a physical realization, we could for instance think
of a long vibrating (``fussy'') polymer.

This paper is organized as follows. In Section \ref{sec:2} we
introduce some of the general theoretical background for the problem.
We then in Section \ref{sec:3}, make our conjecture for the integrated
density of states (IDOS) for our self-affine string based on the
results of Section \ref{sec:2}. In Section \ref{sec:4} we discuss the
decimation technique, the numerical method used to calculated the
IDOS.  The numerical results are presented in Section \ref{sec:5}, and
the conclusion of this paper is drawn in Section \ref{sec:6}.

\section{General Theory}
\label{sec:2}

In this section we review some results which will prove useful in the
later discussion.  Let $\Gamma \subset {\mathbb R}^n$ be a bounded
open set.  Consider the 1D elliptic differential equation
 \begin{eqnarray}
  \label{eq2.1}
 {\bf \nabla^2}u(x)  + a(x)\omega^2 u(x) &=& 0,
 \hspace{1cm} x \in \Gamma
\end{eqnarray}
with homogeneous Dirichlet boundary conditions.  The ``weight
function'' $a(x)$, will be assumed to be a positive real valued
function.  This equation has a countable sequence of positive
eigenvalues (eigenfrequencies) tending to infinity.

Let $N(\omega,\Gamma)$ denote the integrated density of states
(IDOS), i.e.\ the number of eigenmodes with eigenfrequency below
$\omega$. Lapidus and Fleckinger \cite{Fleckinger2} have showed that
the asymptotic behavior of $N(\omega,\Gamma)$ is
\begin{eqnarray}\label{eq2.2} 
   N(\omega,\Gamma) &\sim&
   W(\omega,\Gamma) = \frac{{\cal B}_n \omega^n}{(2\pi)^n}
            \int_\Gamma (a(x))^{n/2} \;d^n\!x,           
\end{eqnarray}
as $\omega \rightarrow \infty$.  Here ${\cal B}_n$ denotes the volume
of the unit ball in ${\mathbb R}^n$ and $\Gamma$ is the resonator domain.
This term is usually called the ``Weyl term'' after Weyl who first
proved this asymptotic behavior for a ``classical'' (i.e.\ non-fractal)
resonator (the {\it Weyl conjecture\/}) \cite{Weyl}.

\section{Conjectures for the self-affine string}
\label{sec:3}

In this paper we will study the situation where the boundary is
regular while the weight-function and/or the elastic coefficient is
irregular.  We will use the following physical model: Consider a
freely vibrating string with fixed endpoints at $x=0$ and $x=L$
(Dirichlet boundary conditions).  It has an elastic coefficient
$E(x)=E_0+E_1(x)$ and density profile $\rho(x)=\rho_0+\rho_1(x)$.
Here $E_0>0$ and $\rho_0>0$ are constants and $E_1(x)$ and
$\rho_1(x)$ are strictly positive, self-affine functions.  The
equation describing the vibrations of this ``rough string'' is
\begin{equation}
\label{eq2.0}
\frac{1}{\rho(x)}\ \frac{d}{dx}\ E(x)\ \frac{d}{dx}\  u(x)\  +\ 
\omega^2\ u(x)\ =\ 0\;.
\end{equation}
If $E(x)=E_0$, then $\rho(x)\equiv a(x)$ in Eq.\ (\ref{eq2.1}) \cite{Graff}. 
   
We will assume that the elastic coefficient $E_1(x)$ and the density
$\rho_1(x)$ vary in a self-affine way \cite{Feder}.  The concept of
self-affinity is a scaling property.  A function defined as $h=h(x)$
is said to self-affine if it is statistically invariant under the
transformation
\begin{mathletters}
\label{eq3.1}
\begin{eqnarray} 
   x   &\rightarrow&  \lambda x\;, \\
   h(x) &\rightarrow&  \lambda^H h(x)\;,
\end{eqnarray}
for all positive $\lambda \in {\mathbb R}$, or equivalently 
\begin{eqnarray} 
   h(x) &\simeq&  \lambda^{-H} h(\lambda x)\;,
\end{eqnarray}
\end{mathletters} 
where $\simeq$ is used in order to indicate statistical equality.  The
parameter $H$ is the Hurst exponent (or roughness exponent).  When
$H>1$, $h(x)$ is not asymptotically flat and the surface is not
statistically invariant under translation.  When $H<0$, the variance
of $h(x)$ diverge when the interval over which it is measures goes to
zero.  $h(x)$ is then referred to as a {\it fractional noise.\/} We
will in our analysis assume $0<H<1$.  Self-affinity is in practice
only found within a restricted range of scales.  In this work we
explicitly introduce a lower cut-off $l$. The upper cut-off is the
system size $L$, which is the length of the string.

Asymptotically, on large scale, $E_1(x)$ and $\rho_1(x)$ will dominate
the behavior of $E(x)$ and $\rho(x)$ no matter what $E_0$ and $\rho_0$
are.  However, we will investigate the system at intermediate scales
where the constant terms may or may not dominate over the self-affine
terms in $E(x)$ and $\rho(x)$.

We will distinguish four cases: 
\begin{enumerate}
  \item $E_0\gg \left<E_1(x)\right>$ and $\rho_0\ll \left<\rho_1(x)\right>$,  
  \item $E_0\ll \left<E_1(x)\right>$ and $\rho_0\ll \left<\rho_1(x)\right>$,
  \item $E_0\ll \left<E_1(x)\right>$ and $\rho_0\gg \left<\rho_1(x)\right>$,
  \item $E_0\gg \left<E_1(x)\right>$ and $\rho_0\gg \left<\rho_1(x)\right>$,
\end{enumerate}
where $\left< \cdot \right>$ denotes the averaging operator.

We note that the left hand side of Eq.\ (\ref{eq2.0}),
$(1/\rho)[(dE_1/dx) (du/dx)+E(d^2u/dx^2)]$, involves the derivative of
a self-affine function, $dE_1/dx$.  This quantity scales as
$dE_1/dx\to \lambda^{H-1} dE_1/dx$ when $x\to\lambda x$, while $E_1$
scales as $E_1\to \lambda^H E_1$.  Thus, we make the assumption that
the term containing $dE_1/dx$ may be neglected in comparison to the
term containing $E_1$ in Eq.\ (\ref{eq2.0}).

With this assumption, we expect the IDOS for the self-affine string to 
behave as ($n=1$, ${\cal B}_1=2$)
\begin{eqnarray} \label{eq3.2} 
   N(\omega,L) &\sim&
   W(\omega,L) = \frac{\omega}{\pi}
            \int_0^L \sqrt{\frac{\rho(x)}{E(x)}} \;d\!x,
            % \hspace{1.2cm} as \;\;\omega \rightarrow \infty
\end{eqnarray} 
in the asymptotic limit.  Note that this expression can be written
$W(\omega,L) = \omega L/\pi\langle\sqrt{\rho(x)/E(x)}\rangle$.
% where $\langle\cdot\rangle$ is the averaging operator.  
Thus, the asymptotic behavior of the IDOS for a rough string is
expected to behave as a classical (non-rough) string with a constant
inverse velocity $\langle\sqrt{\rho(x)/E(x)}\rangle$.

We now discuss the four cases in turn. Assume now that we change the
system size according to $L\rightarrow \lambda L$.  Then, for case (1)
($E_0\gg \left<E_1(x)\right>$ and $\rho_0\ll \left<\rho_1(x)\right>$)
after a change of variable, it follows from Eqs.\ (\ref{eq3.1}) and
(\ref{eq3.2}) that
\begin{mathletters}
 \label{eq3.3a_whole}
\begin{eqnarray}
    \label{eq3.3}
W(\omega,\lambda L) &=& \frac{\omega}{\pi}
            \int_0^{\lambda L} \sqrt{\frac{\rho(x)}{E_0}} \;d\!x \nonumber\\
    &=& \frac{\omega}{\pi} \int_0^{L} \sqrt{\frac{\rho(\lambda x')}{E_0}}
          \;d(\lambda x') \nonumber\\
    &\simeq&  \lambda^{1+\frac{H}{2}}\; W(\omega,L).
\end{eqnarray}
Using Eq.\ (\ref{eq3.2}), this relation may be rewritten
\begin{eqnarray}
    \label{eq3.3a}
    W(\lambda^{-1-\frac{H}{2}}\,\omega,\lambda L)&\simeq& W(\omega,L).
\end{eqnarray}
\end{mathletters}
Note that Eqs.\ (\ref{eq3.3}) and (\ref{eq3.3a}) 
are in principle equivalent, 
but open up two different possibilities for physical interpretation. 
In the former case, it is the number of eigenmodes which is scaled,
while in the latter it is the frequency.  

Through similar arguments, we find for case (2) 
($E_0\ll \left<E_1(x)\right>$ and $\rho_0\ll\left<\rho_1(x)\right>$),
\begin{mathletters}
\label{eq3.3b_whole}
\begin{equation}
  \label{eq3.3b}
  W(\omega,\lambda L) \simeq \lambda W(\omega,L)\;,
\end{equation}
or equivalently
\begin{equation}
\label{eq3.3c}
  W(\lambda^{-1}\omega,\lambda L) \simeq W(\omega,L)\;.
\end{equation}
\end{mathletters}
For case (3) ($E_0\ll\left<E_1(x)\right>$ and $\rho_0\gg
\left<\rho_1(x)\right>$), we expect
\begin{mathletters}
\label{eq3.3d_whole}
\begin{eqnarray}
  \label{eq3.3d}
  W(\omega,\lambda L) \simeq \lambda^{1-H/2} W(\omega,L)\;, \\ 
%\end{equation}
% or
%\begin{equation}
  \label{eq3.3e}
  W(\lambda^{-1+H/2}\omega,\lambda L) \simeq W(\omega,L)\;.
\end{eqnarray}
\end{mathletters}
Case (4) ($E_0\gg \left<E_1(x)\right>$ and $\rho_0\gg\left<\rho_1(x)\right>$) leads to the same
behavior as Case (2), Eqs.\ (\ref{eq3.3b_whole}).

\section{Outline of the numerical method}
\label{sec:4}
                       
In order to do numerical simulation of our self-affine string
we discretize Eq.\ (\ref{eq2.0}) on a lattice of $N$ sites
\begin{eqnarray} \label{eq4.1}
    A^{(0)}_{ij} U_j(\omega) &=& \omega^2 U_i(\omega)
      \hspace{1cm} i,j=1,2,\ldots,N.
\end{eqnarray}
Here $A^{(0)}$ is the $N\times N$ matrix representation for the operator 
$(1/\rho(x))(d/dx)E(d/dx)$.  It is tridiagonal.
 
The method used in this paper to calculate the IDOS for the above
equation is the decimation technique of Lambert and Weaire \cite{Lambert}.
This method is based on a renormalization philosophy, 
where successive degrees of freedom are eliminated from the system,
see Refs.\ \cite{Lambert,Burton} for more details.
After removing the sites corresponding to $k=N-M,\ldots,N$ the
system becomes~\cite{Williams}
\begin{eqnarray} \label{eq4.2}
      A^{(M)}_{ij} U_j(\omega) &=& \omega^2 U_i(\omega),
      \hspace{1cm} i=1,\ldots,N-M
\end{eqnarray}
with
\begin{eqnarray} \label{eq4.3}
   A^{(M)}_{ij} &=& A^{(M-1)}_{ij} +
        \frac{A^{(M-1)}_{i,N+1-M} A^{(M-1)}_{N+1-M,j}}{D^{(M)}}
\end{eqnarray}
where the denominator is given by
\begin{eqnarray}
 D^{(M)} &=&  A^{(M)}_{N+1-M,N+1-M} - \omega^2\;.
\end{eqnarray}

At the very heart of the method lies the fact that 
renormalized system is equivalent to the original one in the sense
that the two systems have the same eigenvalues. By repeating this
procedure all degrees of freedom can be eliminated.
The number of eigenmodes less then $\omega$, i.e. the IDOS,
can now be shown to be equal to the number of 
negative denominators in the sequence
$\{D^{(i)}\}_{i=1}^N$~\cite{Tremblay}.
We would like to point out that this algorithm is very efficient.

\section{Simulation and results}
\label{sec:5}

In order to test the conjecture\ (\ref{eq3.3a_whole}), and thus also
Eqs.\ (\ref{eq3.3b_whole}) to (\ref{eq3.3d_whole}), we have calculated
the IDOS for various system sizes, $L$, fixing the Hurst exponent $H$
to the value $0.7$.  We work in length units where $l=1/2^{11}$. Thus,
the system size $N=2^{11}=2048$ corresponds to a string of unit length
($L=1$).  We set $E_0=1$ and $\left<E_1(x)\right>=0.01$ for the cases
where $E_0\gg \left<E_1(x)\right>$, and $E_0=0.01$ and
$\left<E_1(x)\right>=1$ for the cases where $E_0\ll
\left<E_1(x)\right>$. Likewise, we set $\rho_0=1$ and
$\left<\rho_1(x)\right>=0.01$ for the cases where $\rho_0\gg
\left<\rho_1(x)\right>$, and $\rho_0=0.01$ and
$\left<\rho_1(x)\right>=1$ for the cases where $\rho_0\ll
\left<\rho_1(x)\right>$.  We averaged over 500 samples in case (1),
while we used 50 samples in the three other cases.

\begin{figure}
    \begin{center}
        \begin{tabular}{@{}c@{\hspace{1.0cm}}c@{}}
            \epsfig{file=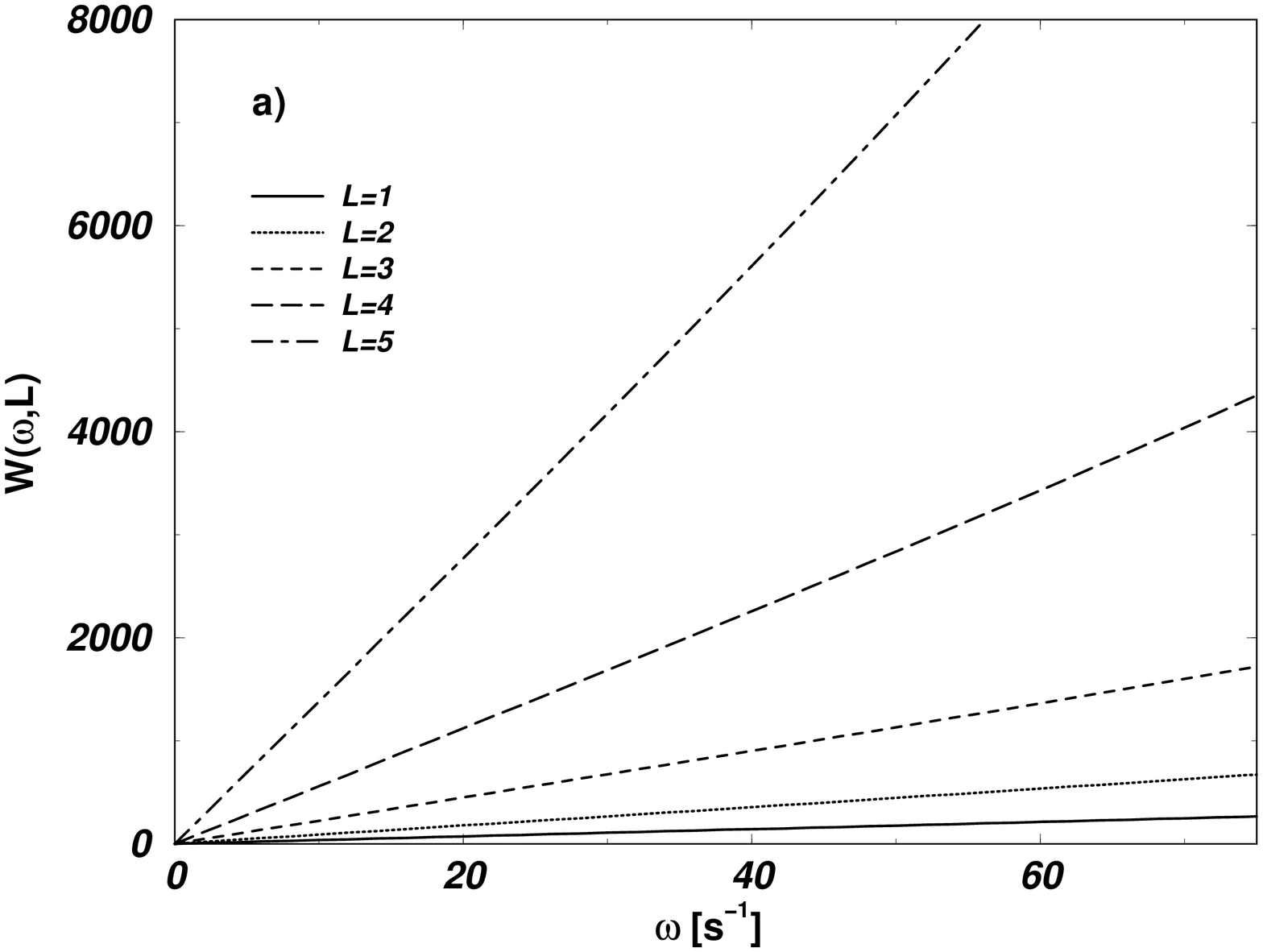,width=7.5cm,height=7.5cm,angle=-90} &
            \epsfig{file=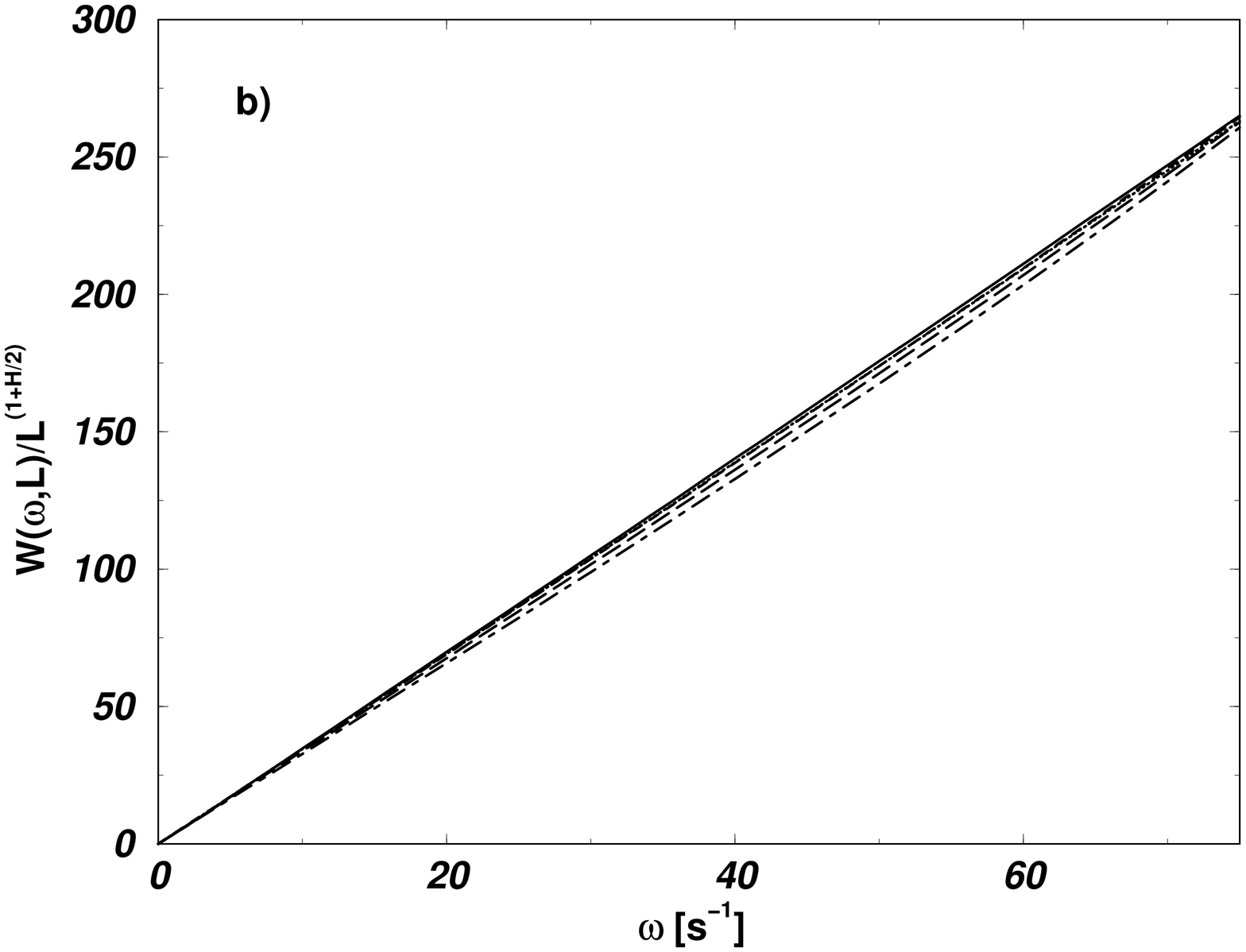,width=7.5cm,height=7.5cm,angle=-90} 
        \end{tabular}
    \end{center}
    \caption{(a) The asymptotic behavior of the integrated density of
      states (IDOS) when $E_0\gg \left<E_1(x)\right>$ and $\rho_0\ll
      \left<\rho_1(x)\right>$ (Case 1).  (b) The scaled IDOS
      $W(\omega,L)/L^{1+H/2}$ as function of $\omega$.}
    \label{fig1}
\end{figure}

As can be seen from Figs.\ \ref{fig1}a the IDOS is a linear function
of the frequency $\omega$, and $W(\omega,L)$ increases with system
size $L$. This is consistent with the predictions of Eq.\ 
(\ref{eq3.3}). A similar linear behavior has been found for case (2)
to (4), but these results are not shown explicitly.

\begin{figure}
    \begin{center}
        \begin{tabular}{@{}c@{\hspace{1.0cm}}c@{}}
            \epsfig{file=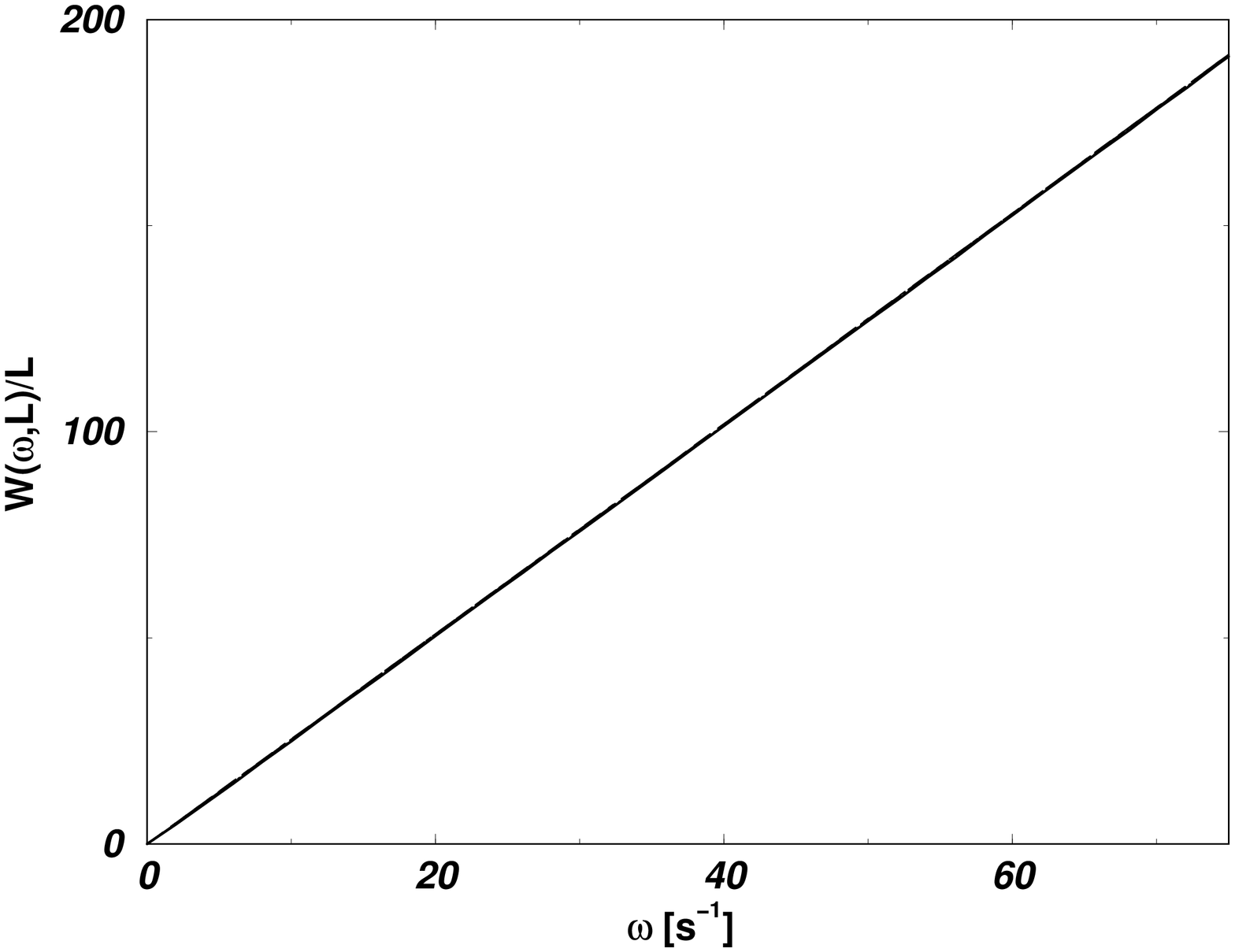,width=7.5cm,height=7.5cm,angle=-90}
        \end{tabular}
    \end{center}
    \caption{The scaled IDOS $W(\omega,L)/L$
      as function of $\omega$ for Case (2): $E_0\ll
      \left<E_1(x)\right>$ and $\rho_0\ll \left<\rho_1(x)\right>$.}
    \label{fig2}
\end{figure}

\begin{figure}
    \begin{center}
        \begin{tabular}{@{}c@{\hspace{1.0cm}}c@{}}
            \epsfig{file=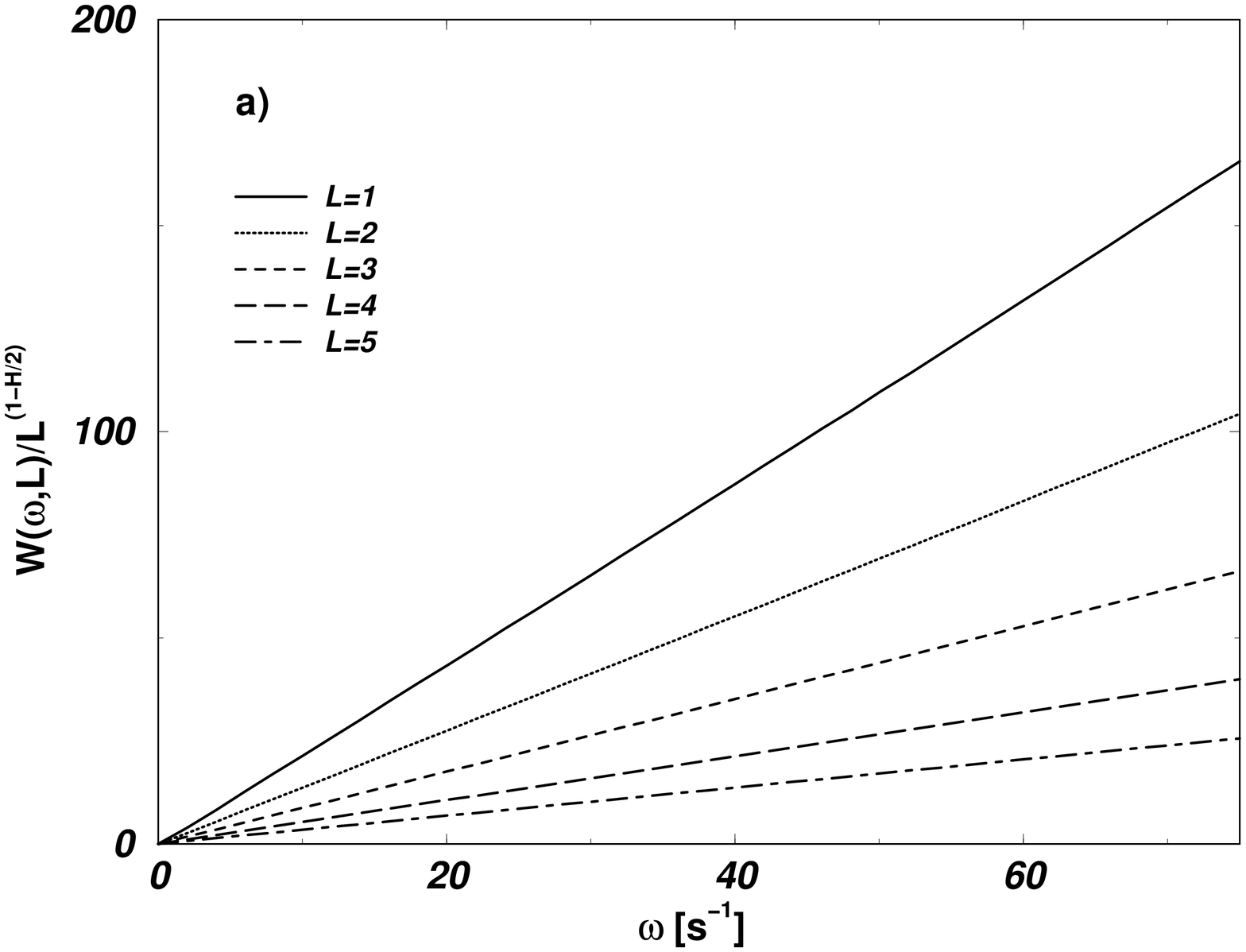,width=7.5cm,height=7.5cm,angle=-90} &
            \epsfig{file=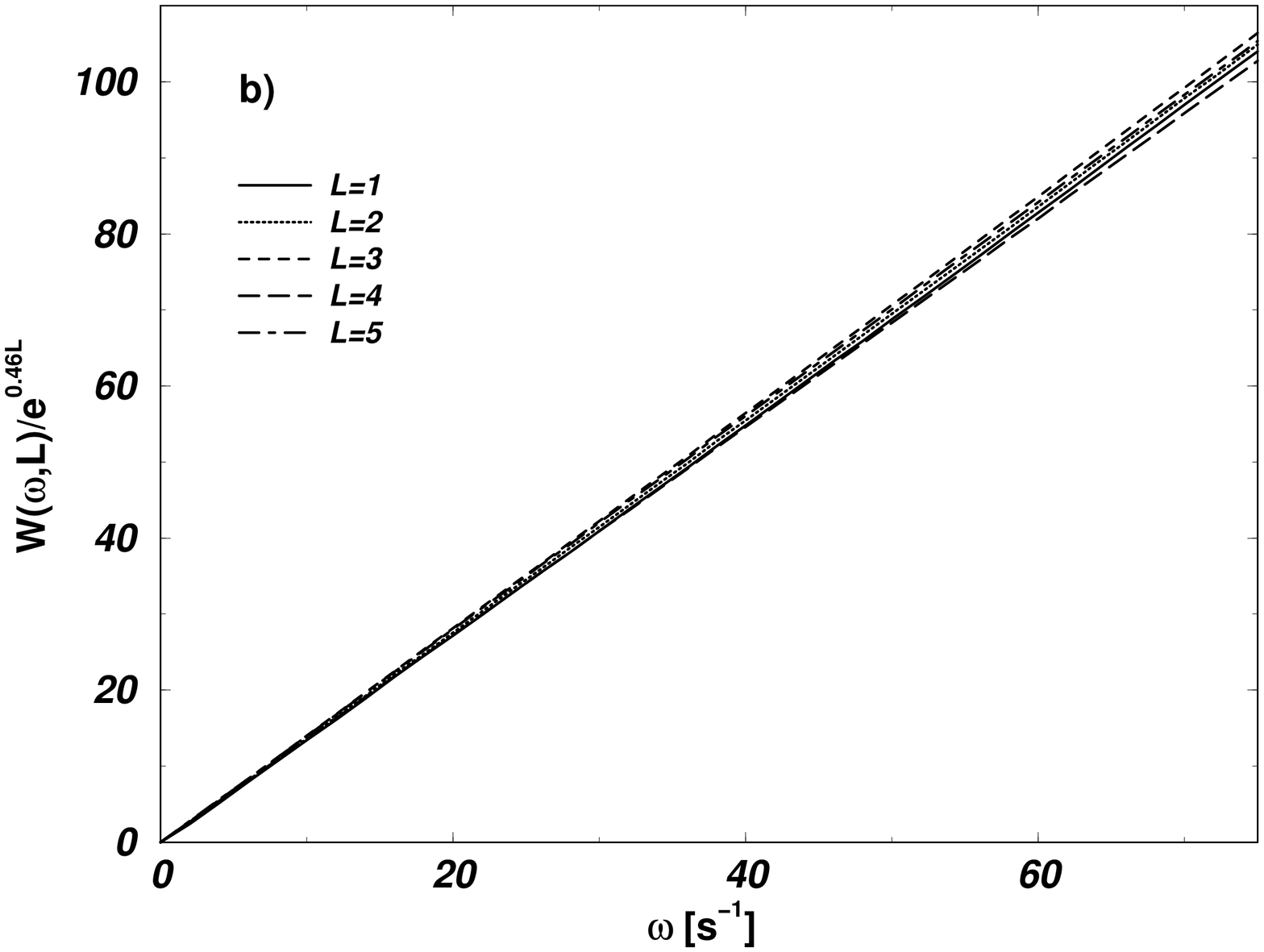,width=7.5cm,height=7.5cm,angle=-90} 
        \end{tabular}
    \end{center}
    \caption{The scaled IDOS (a)~$W(\omega,L)/L^{1-H/2}$ and  
      (b)~$W(\omega,L)/e^{0.46L}$ as function of $\omega$ for Case
      (3): $E_0\ll \left<E_1(x)\right>$ and $\rho_0\gg
      \left<\rho_1(x)\right>$.}
    \label{fig3}
\end{figure}

\begin{figure}
    \begin{center}
        \begin{tabular}{@{}c@{\hspace{1.0cm}}c@{}}
            \epsfig{file=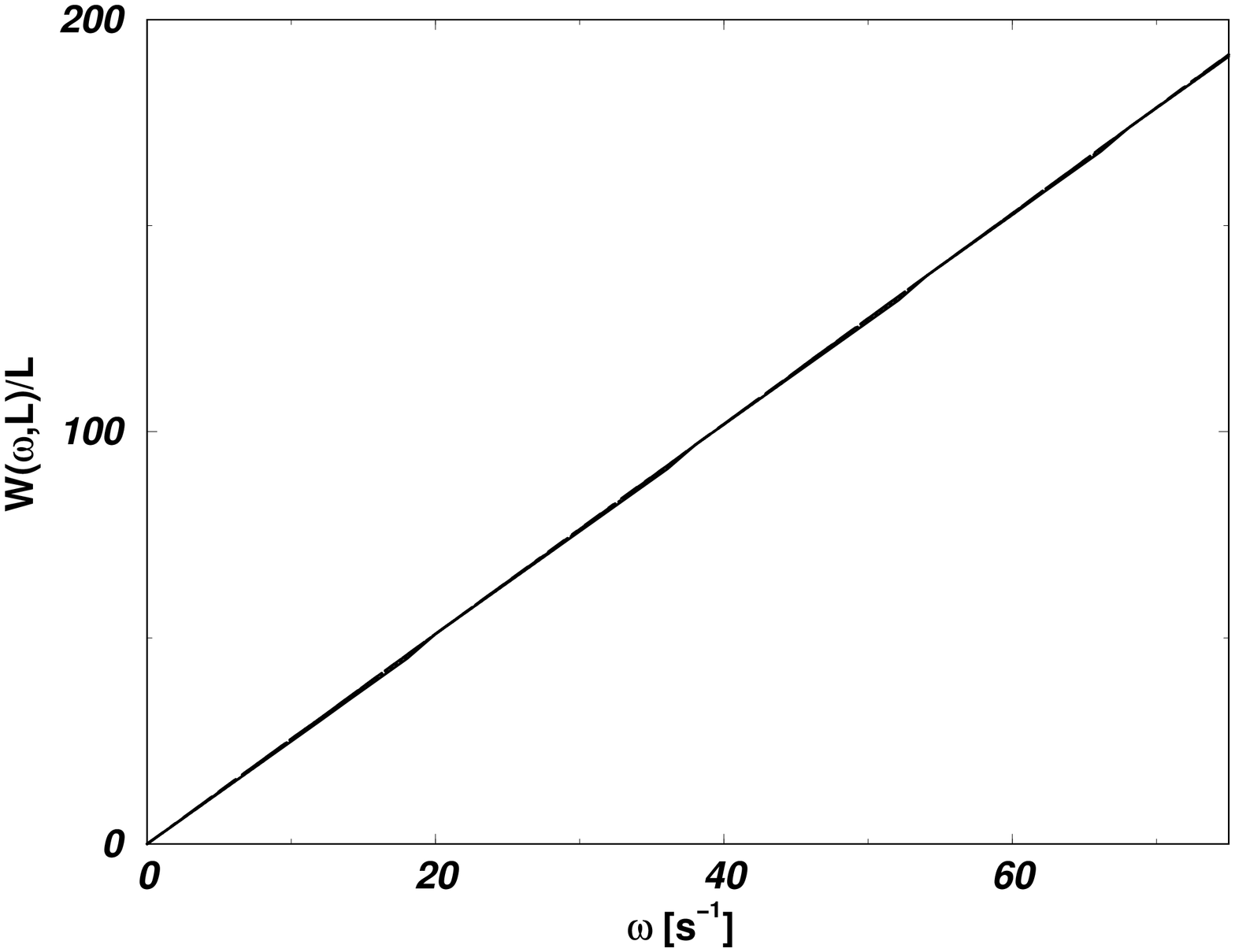,width=7.5cm,height=7.5cm,angle=-90}
        \end{tabular}
    \end{center}
    \caption{The scaled IDOS $W(\omega,L)/L$
      as function of $\omega$ for Case (4): $E_0\gg
      \left<E_1(x)\right>$ and $\rho_0\gg \left<\rho_1(x)\right>$.}
    \label{fig4}
\end{figure}

We test in Fig.\ \ref{fig1}b the case (1) scaling relation (\ref{eq3.3}),
in Fig.\ \ref{fig2} the case (2) scaling relation (\ref{eq3.3b}),
in Fig.\ \ref{fig3}a the case (3) scaling relation (\ref{eq3.3d}), and
in Fig.\ \ref{fig4} the case (4) scaling relation (\ref{eq3.3b}).

In all cases {\it except\/} case (3), our numerical data are
consistent with the theoretical predictions.  In case (3), where
$E_0\ll \left<E_1(x)\right>$ and $\rho_0\gg \left<\rho_1(x)\right>$,
data collapse was obtained by scaling $W(\omega,L)/\exp(0.46 L)$
(see Fig.~\ref{fig3}b).  This is very different from the predicted
behavior, Eq.\ (\ref{eq3.3d}).  We do not know the reason for this
discrepancy between theory and numerical result.

\section{Summary and conclusion}
\label{sec:6}
 
We have investigated the IDOS for a self-affine string, i.e.\ a string
with self-affine variations in mass density and elastic coefficient.
There are four relevant cases to study depending on whether the
self-affine variations dominate or not in the mass density and elastic
coefficient.  We have compared numerical studies with a conjecture
based on assuming that a result of Lapidus and Fleckinger
\cite{Fleckinger2} is valid for the self-affine string.

We find that our conjecture works for three out of the four cases.
However, it fails for the case where $E_0\ll \left<E_1(x)\right>$ and
$\rho_0\gg \left<\rho_1(x)\right>$.

\acknowledgements

One of the authors (I.S.) would like to thank the Research Council of
Norway and Norsk Hydro ASA for financial support. A.H.\ would like to
thank H.N.\ Nazareno and F.A.\ Oliveira for warm hospitality and the
I.C.C.M.P.\ for support.  This work has received support from the
Research Council of Norway (Program for Supercomputing) through a
grant of computing time.

% --------------------------------------------------------------------
% BIBLIOGRAPHY
% --------------------------------------------------------------------

% --------------------------------------------------------------------
% FIGURE CAPTIONS
% --------------------------------------------------------------------

\end{multicols}
\widetext
\newpage

\end{document}